\documentclass[aps,pra,amsmath,amssymb, preprint,superscriptaddress, floatfix]{revtex4-2}
\usepackage{amssymb}
\usepackage{amsmath}
\newcommand{\bol}[1]{\boldsymbol{#1}}
\usepackage{physics}
\usepackage{calc}
\usepackage{mathtools}
\usepackage{enumitem}
\usepackage[pdftex,pdfpagelabels,bookmarks,hyperindex,hyperfigures,colorlinks]{hyperref}
\begin{document}

%\preprint{}

\title{Space Time Algebra Formulation of Cold Magnetized Plasmas}


\author{Kyriakos Hizanidis}
\email{kyriakos@central.ntua.gr}
\affiliation{School of Electrical and Computer Engineering, National Technical University of Athens, Zographou 15780, Greece}
\author{Efstratios Koukoutsis}
\affiliation{School of Electrical and Computer Engineering, National Technical University of Athens, Zographou 15780, Greece}
\author{Panagiotis Papagiannis}
\affiliation{School of Electrical and Computer Engineering, National Technical University of Athens, Zographou 15780, Greece}
\author{Abhay K. Ram}
\affiliation{Plasma Science and Fusion Center, Massachusetts Institute of Technology, Cambridge, Massachusetts 02139, USA}
\author{George Vahala}
\affiliation{Department of Physics, William \& Mary, Williamsburg, Virginia 23187, USA}

\date{\today}

\begin{abstract}
The propagation and scattering of electromagnetic waves in magnetized plasmas in a state where a global mode has been established or is in turbulence, are of theoretical and experimental interest in thermonuclear fusion research. Interpreting experimental results, as well as predicting plasma behavior requires the numerical solutions of the underlying physics, that is, the numerical solution of Maxwell equations under various initial conditions and, under the circumstances, complex boundary conditions.  Casting, the underlying equations in a coordinate free form that exploits the 
symmetries and the conserved quantities in a form that can easily encompass a variety of initial and boundary conditions is of tantamount importance. Pursuing this task we utilize the advantages the Clifford Algebras can possibly provide. For simplicity we deal with a cold multi-species lossless magnetized plasma. The formulation renders a Dirac type evolution equation for am augmented state that consists of the electric and magnetic field bivectors as well as the polarizations and their associated currents for each species. This evolution equation can be dealt with a general spatial lattice disretization scheme. The evolution operator that dictates the temporal advancement of the state is Hermitian. This formulation is computationally simpler whatever the application could be. However, small wavelength capabilities (on the Debye length scale) for spatially large systems (magnetic confinement devices) is questionable even for conventional super-computers. However, the formulation provided in this work it is entirely suitable and it can be directly transferred in a quantum computer. It is shown that the simplified problem in the present work could be suitable for contemporary rudimentary quantum computers.
\end{abstract}

\maketitle

% body of paper here -Use  proper section commands
% References should be done using the \cite, \eqref, and \label commands
\section{Introduction}\label{Introduction}
Electromagnetic waves are abundant in Nature and in the laboratory playing a useful role in a variety of applications ranging from communications to heating of thermonuclear fusion plasma. In the latter, the propagation and scattering of electromagnetic waves and wave-packets in the edge and in the main bulk of the magnetically confined plasmas play a significant role for their confinement capability. The full problem by itself poses an arduous analytical and computational task because of the ambiguities of the initial conditions and, most importantly, because of the complexity in the boundary conditions and the geometric characteristics of the plasma itself. However, geometrically symmetric settings are possibly amenable to analytical, semi-analytical treatments and they are more simple computationally. Recently, full-wave theoretical models, using Maxwell’s equations,has been developed based on the Mie theory for scattering of electromagnetic waves by spherical blobs and cylindrical density filaments \cite{Ram1,Ram2,Ram3}. These models have been applied using computational models implemented on classical computers. These models utilize the series representations of vector cylinder functions \cite{Stratton} or vector spherical harmonics \cite{Barrera}. Thus, an efficient and straightforward computer-ended analytical solutions of the scattering problem emanates as compared to other numerical schemes, such as, finite elements methods. However, the obtained solutions (by matching the series expansions on the boundaries of these formations) necessarily involve series truncation that depend on the wavelength of the incident beam and the size of the blobs or the density fluctuations. Thus, obtaining sufficient accuracy is a delicate issue that can only be resolved on a case-by-case approach. 

For more general geometrical settings, one must relay on conventional computer-intensive methods in solving Maxwell equations or the vectorial Helmholtz equation under given initial and boundary conditions. Alternatively, one can envision a more general spatial discretization (a lattice scheme, in other words) combined with a coordinate-free solutions to Maxwell equations. In this scheme the first order spatial derivatives involved can be readily and simply represented. Thus, the ideal scheme could be solving Maxwell equations as an evolution problem (first order in time) of a state that will represent the electromagnetic fields. In a first approximation, if one restricts oneself to a linear material response, this task would be easier to tackle. In the present work, we focus on the geometric representation of the electromagnetic field considering the magnetically confined plasma as a cold dissipation-free gyrotropic dielectric medium. Thus, one expects conservation laws to hold (electromagnetic energy, power flow, etc) and, therefore, the evolution equation we are looking for, must incorporate these laws. Our task is to recast the electromagnetic propagation and scattering problem in cold magnetized plasmas in a form suitable for evolution of a lattice scheme.  Geometric algebras (GAs) offer an elegant way to provide a coordinate free description of the electromagnetism \cite{Dressel}. 

The question that naturally arises is if elegance is a high price to pay in adapting the geometric approach. However, it is generally accepted in the scientific community that GAs are, by far, the easiest way to describe rotations in space and in higher dimensions \cite{Doran}. This becomes more obvious if one deals with such transformations in the realm of special relativity and the associated Lorentz transformations [fundamental to electromagnetism (EM)] in four dimensional Minkowskian spacetime. Note that, the application of GAs always leads to a change of perspective.  However, this ingenious approach has been introduced many decades ago and found a fruitful field of application in other branches of technological endeavor such as computer vision and robotics, multi-component chemical compositions as well as and multi-particle quantum systems \cite{Dorst,Doran,Arthur}. To our knowledge the proposed application for such a medium (gyrotropic) is new in the literature. In a recent work of ours, \cite{Koukoutsis2} the problem of the evolution of the EM field in a cold magnetized gas of various charged particles has been formulated as an initial value problem: In this approach, the initial state of the system is the vacuum state. Thus, the medium is not initially polarized. In the present work, we implicitly assume that the plasma is already in a fully developed state, e.g., some global mode, or there exist a background EM spectrum and/or turbulent fluctuations. To these extent, we expect that the mixture of the preexisting polarizations will play a very important role, in contrast with \cite{Koukoutsis2}.

An additional advantage of the proposed method is that it can be readily transferred and adapted to the so-called Qubit Lattice Agorithms (QLA) \cite{Koukoutsis,Vahala1,Ram,Vahala2}. These algorithms are basically meant for quantum computers and quantum encoding, a rapidly advancing field in the realm of Quantum Information and implementations via Quantum Computing (QC) techniques \cite{Nielsen}. The reason is that GAs offer an elegant way to describe rotations in Bloch hyper-spheres that are synonymous to the unitary operators acting on qubit states. Nevertheless, the expected outcome of the present work, namely the lattice evolution equation can be readily implemented on state-of-the-art conventional computers. The computational time required to solve such evolution equations will eventually depend on the spatial complexity of our scattering problem. However, even with these limitations, the scattering and propagation problem will be less demanding computationally. Note, also that, due to the straightforward portability (and, most importantly, linearity) of the proposed method for future quantum implementation, these limitations could eventually be overcome.

In Sec.\eqref{Dirac form of Maxwell equations in cold magnetized plasmas}, we first introduce the basic element that carries all the information for the electromagnetic field. Then, we formulate the Dirac-Schr\"odinger-type equation that follows the evolution of the latter through the plasma along with the linear plasma response. The latter introduces auxiliary field entities which are related to the wave polarization and to the polarization density current. These auxiliary entities lead to an augmented state description. In the last Sec.\eqref{Conclusion} the main results are summarized. Additionally, in Sec.\eqref{Appendix A} a brief introduction on the basic GA mathematical tools used thoughout this work is presented and in 
Sec.\eqref{Appendix B} a rudimentary estimation on the necessary resources for classical and quantum implementation is provided by an example from today's thermonuclear fusion endeavor. 

\section{Dirac form of Maxwell equations in cold magnetized plasmas}\label{Dirac form of Maxwell equations in cold magnetized plasmas}
A perspective of Maxwell equations that will both serve short and long term implementations is to recast these equations into a form that is similar to the Dirac or Schr\"odinger equations of quantum mechanics for closed quantum systems. Classical, linear and energy conserving systems admit a Schr\"odinger-type representation analogous to the unitary evolution (corresponding to energy conservation) of quantum systems. In the realm of the geometric representation of the electromagnetic field in hand, there exist two geometric approaches \cite{Dressel}: Either representing the field as a mixture of a vector and a bivector, frequently called biparavector, an alternative to the extensively used Riemann-Silberstein-Weber vector (RSW) \cite{Silberstein}, or as a genuine bivector, $\mathcal{F}$. The geometric algebra behind the first approach is the Clifford Geometric Algebra $\mathcal{C}l({\mathbb{R}}^{3})$ in the three-dimensional space, ${\mathbb{R}}^{3}$, which is called Pauli Algebra $\mathcal{PA}$. The respective geometric algebra behind the second approach is, instead, the Clifford Geometric Algebra $\mathcal{C}l({\mathbb{R}}^{1,3})$ which is called Dirac Algebra, $\mathcal{DA}$, or, alternatively, spacetime geometric algebra (STGA)], which refers to the four dimensional Minkowskian spacetime, ${\mathbb{R}}^{1,3}$, with the metric $(1,-1,-1,-1)$. The underlying correspondence between this  approach and Maxwell equations has been the subject of early work by Vaz and Rodrigues (\cite{Vaz1}, \cite{Vaz2}).  

In this work we adopt the second approach as far more suitable since it brings in naturally Lorentz invariance and symmetries of the EM. Besides, this description treats the electromagnetic field as real, that is, the amplitudes of all bivector involved are real quantities. This allows the description to be free of phase redundancies that are common in RSW descriptions or more classical ones that use phasors. Note that $\mathcal{DA}$ is isomorphic to the algebra of $4X4$ real matrices (while $\mathcal{PA}$ is isomorphic to the algebra of $2X2$ complex matrices).  The EM bivector $\mathcal{F}$ consists of six bivectorial components (see in \eqref{sec:3}): The first three correspond to the Gibbsian (i.e. vectorial in the Cartesian sense) electric field intensity, $\boldsymbol{E}$ and they are time-like, that is, there are defined by the three blades (geometric product of mutually orthogonal directions in spacetime) that possess time-like directions, i.e., the geometric products $\gamma_m\gamma_0,(m=1,2,3)$  with $\gamma_0$ being the unit vector along the temporal direction ($0$). The latter three components are purely spatial left-handed (32, 13, 21) unit blades,  $\gamma_m\gamma_n,(m,n=1,2,3)$ and correspond to the Gibbsian magnetic field induction, $\boldsymbol{B}$. In the following, the script capital letter signifies a bivector which usually contains space-time ($\gamma_m\gamma_0$) partial bivectors (space-time blades) and space-like bivectors ($\gamma_m\gamma_n, m=1,2,3$). Adapting this notation, the EM bivector is defined (along with its Clifford conjugate, obtained in Sec.\eqref{Appendix A}, \eqref{EM bivector properties}) as follows:
\begin{equation}\label{EM bivector 1}
\mathcal{F}=\mathcal{E} +c\mathcal{I}\mathcal{B},\quad  \mathcal{F}^\dagger=\mathcal{E} -c\mathcal{I}\mathcal{B},
\end{equation}
where $\mathcal{E}$ and $\mathcal{B}$ are clearly space-time bivectors (the electric field intensity and the magnetic induction, respectively) and $\mathcal{I}$, which left-multiplies the respective bivectors, is the pseudo-scalar element of $\mathcal{DA}$,
\begin{equation}\label{pseudoscalar}
\mathcal{I}=\gamma_{0123}=\gamma_0\gamma_1\gamma_2\gamma_3,
\end{equation}
In Cartesian component form (summation convention is adopted) one may write:
\begin{equation}\label{components}
\mathcal{E}=E_m\gamma_m\gamma_0,\quad\mathcal{B}=B_m\gamma_m\gamma_0
\end{equation}
The scalar amplitudes $E_m,B_m$, obviously (\eqref{orthonormality}) are:
\begin{equation}\label{amplitudes}
E_m=\gamma_0\bol{\cdot}\mathcal{E}\bol{\cdot}\gamma_m,\quad B_m=\gamma_0\bol{\cdot}\mathcal{B}\bol{\cdot}\gamma_m
\end{equation}
The EM bivector and its Clifford conjugate bears the property of a sign change upon reversion (defined in Sec.\eqref{Appendix A}), that is:
\begin{equation}\label{EM reversed}
\tilde{\mathcal{F}}=-\mathcal{F}, \quad \tilde{\mathcal{F}^\dagger}=-\mathcal{F}^\dagger\end{equation}
Another important property of the EM bivector is that its symmetrized geometric product with its Clifford conjugate is scalar, that is, of grade zero. We use the term $\emph{dot product}$ for this scalar part and use the respective symbol. It is straightforward to see that this product is directly realated to the EM energy density in the vacuum,$E_0$, (summation convention is adapted):
\begin{equation}\label{EM energy}
\mathcal{F}\bol{\cdot}\mathcal{F}^\dagger\equiv \frac{\mathcal{F}\mathcal{F}^\dagger + \mathcal{F}^\dagger\mathcal{F}}{2}= E_mE_m+c^2B_mB_m=|\bol{E}|^2+c^2|\bol{B}|^2 =\frac{2E_0}{\epsilon_0}
\end{equation}
where $\bol{E}$ and  $\bol{B}$ are the respective Cartesian vectors for the electric field intensity and the magnetic induction.

The first three components in the bivector expressions in \eqref{EM bivector 1} are time-like, that is, there are defined by the three blades that possess time-like directions, i.e., $\{\gamma_m\gamma_0, m=1,2,3\}$ and correspond to the electric field intensity. The other three components are purely space-like, i.e. $\{\gamma_m\gamma_n, m,n=1,2,3\}$ and the corresponding space-like unit blades are left-handed $(32, 13, 21)$ and correspond to the magnetic field induction. However, the latter can also be considered as the result of the duality transformation (\eqref{dual}) of a time-like object, like the electric field intensity.  Thus, the magnetic field induction will be the result of the geometric product of the pseudoscalar of $\mathcal{DA}$ and this time-like object [multiplied by the speed of light in the vacuum, $c$, to render it in the same MKSA units $(V/m)]$. Note also (with respect to the EM bivector):
\begin{equation}\label{E and B bivectors}
\mathcal{E}=\frac{\mathcal{F}^\dagger+\mathcal{F}}{2},\quad c\mathcal{B}=\mathcal{I}\frac{\mathcal{F}^\dagger-\mathcal{F}}{2}
\end{equation}

We now proceed with the geometric prospective of the differential laws, connected with the well-known Maxwell equations in Ampere's formulation: The latter formulation is better adapted for dielectric media, such as plasmas, where the bound currents and electric charges due to the applied fields can be incorporated as parts of the sources, especially for the case (our case) where free currents and charges are absent. The geometric perspective must possesses the following capabilities: (1) Contraction-grade reduction of the differential properties of the EM bivector $\mathcal{F}$ that finally amounts to the well-known divergence laws. (2) Expansion-grade increase of these differential properties that will eventually lead to the fulfillment of the Bianchi identity (in tensorian language). The latter is a requirement in Classical Electromagnetism since it excludes space-time handedness. All these can be incorporated if one considers the differential operator $\mathcal{D}$ operating geometrically on the EM bivector $\mathcal{F}$, that is, in accordance with \eqref{geometric product 2}:
\begin{equation}\label{DF}
\mathcal{D}\mathcal{F}=\mathcal{D}\bol{\cdot}\mathcal{F}+\mathcal{D}\wedge\mathcal{F}
\end{equation}
In this expression, the products (inner and wedge) are respectively connected with the grade reduction and increase. The latter two are purely mathematical processes. Nevertheless they posses clear physical meaning which is related to the two geometrical features of the electromagnetic field considered as a Cartesian vector field: Namely, field line convergence and rotation in the 3-dimensional space. This expression is set equal to the sources. In our case the sources are the induced polarization charge, a scalar function $\rho_b$ and the space-time bivector current function (the subscript "b" stands for "bound"): $\mathcal{J}_b$:
\begin{equation}\label{Maxwell}
\mathcal{D}\mathcal{F}={1\over\epsilon_0}(\rho_b+\mathcal{J}_b/c)\gamma_0
\end{equation}
For dielectric media the functions involved in the RHS of this equation are related to the polarization space-time bivector $\mathcal{P}$.  In Cartesian component form, the latter is (summation convention is adapted):
\begin{equation}\label{polarization components}
\mathcal{P}=P_m\gamma_m\gamma_0
\end{equation}
while in the Ampere's formulation one obtains: 
\begin{equation}\label{polarization charges and currents}
\rho_b = \gamma_0\overleftrightarrow{\bol{\nabla}}\bol{\cdot}\mathcal{P}\gamma_0=-\partial^m P_m,\quad \mathcal{J}_b=c\partial^0\mathcal{P},
\end{equation}
Note that, from the expression \eqref{polarization charges and currents}, it is clear that the polarization currents and charges are tightly connected through the polarization charge conservation law:
\begin{equation}\label{polarization charge conservation}
c\partial^0\rho_b+\partial^m\gamma_0\bol{\cdot}\mathcal{J}_{b}\bol{\cdot}\gamma_m=0
\end{equation}
It straightforward to re-express \eqref{Maxwell} (see Sec.\eqref{Appendix A}) as follows:
\begin{equation}\label{Maxwell again}
\mathcal{D}\mathcal{F}=-{1\over\epsilon_0}\mathcal{D}^\dagger\bol{\cdot}\mathcal{P}
\end{equation}

The space-time bivector $\mathcal{P}$ is related, exclusively, to the electric field intensity bivector. Thus, it is related, according to \eqref{E and B bivectors}, to both $\mathcal{F}$ and its Clifford conjugate (thus, it is identical to its Clifford conjugate). Therefore, one needs to couple \eqref{Maxwell} to its Clifford conjugate:
\begin{equation}\label{conjugate Maxwell again}
(\mathcal{D}\mathcal{F})^\dagger=-{1\over\epsilon_0}(\mathcal{D}^\dagger\bol{\cdot}\mathcal{P})^\dagger
\end{equation}
Via Equations \eqref{conjugate of the reverse of a dot product} and \eqref{conjugate of DF} (see Sec.\eqref{Appendix A}) the last equation yields the second of the following coupled equations:
\begin{equation}\label{Maxwell 2}
\mathcal{D}\mathcal{F}=-{1\over\epsilon_0}\mathcal{D}^\dagger\bol{\cdot}\mathcal{P},\quad \mathcal{D}^\dagger\mathcal{F}^\dagger= -{1\over\epsilon_0}\mathcal{D}\bol{\cdot}\mathcal{P}
\end{equation}
For a multi-species plasma:
\begin{equation}\label{multi-species}\mathcal{P}\equiv\Sigma_\alpha\mathcal{P}^{(\alpha)}\equiv\Sigma_\alpha\mathcal{P}^{(\alpha)}
\end{equation}
The two equations in \eqref{Maxwell 2} can be combined into a compact form as follows:
\begin{equation}\label{matrix form 1}
\begin{bmatrix}
\mathcal{D}&0\\
0&\mathcal{D}^\dagger
\end{bmatrix}\begin{bmatrix}\mathcal{F}\\ \mathcal{F}^\dagger\end{bmatrix}=-{1\over\epsilon_0}\begin{bmatrix}
\mathcal{D}^\dagger&0\\
0&\mathcal{D}
\end{bmatrix}\bol{\cdot}\Sigma_\alpha\begin{bmatrix}\mathcal{P}^{(\alpha)}\\ \mathcal{P}^{(\alpha)}\end{bmatrix}\end{equation}
Note that this compact form is for two \emph{independent} equations that couple the evolution of the EM field to the polarization bivector $\mathcal{P}$.
Equations \eqref{matrix form 1} incorporate all four Maxwell equations that apply for the dielectric medium at hand. It is going to be exploited for the case of a homogeneous cold magnetized plasma. For a locally homogeneous, cold magnetized plasma in the linear response regime, the Cartesian components of the polarization are (summation convention is adapted) are:
\begin{equation}\label{linear polarization 1}
P_m^{(\alpha)}(\{\chi_p\},\chi_0)={\epsilon_0\over c}\int_0^\infty\kappa^{(\alpha)}_{mn}(\{\chi_p\},\xi_0)E_n(\{\chi_p\},\chi_0-\xi_0)d\xi_0
\end{equation}
In these expression $\{\chi_p\}$ refers to a spatial position ($p=1,2,3$) while the superscript $\alpha$ labels a particular species (electron or ion). The temporal variable $\chi_0$ is in spatial units (via the speed of light in the vacuum, $c$). The convolution integral in \eqref{linear polarization 1} contains the matrix of the cold magnetized plasma susceptibility temporal kernel for that particular species (in $s^{-1}$), $\kappa^{(\alpha)}_{mn}(\{\chi_p\},\xi_0)$. From \eqref{amplitudes} and \eqref{E and B bivectors} one alternatively obtains (partially suppressing, for simplicity, the spatio-temporal dependence of the quantities involved):
\begin{equation}\label{linear polarization 2}
P_m^{(\alpha)}(\mathcal{F},\mathcal{F}^\dagger)={\epsilon_0\over c}\int_0^\infty\kappa^{(\alpha)}_{mn}(\gamma_0\bol{\cdot}\mathcal{E}\bol{\cdot}\gamma_n  )_{\chi_0-\xi_0}d\xi_0
\end{equation}
On the other hand the space-time bivector $\mathcal{P}^{(\alpha)}$ can also be written as (double summation convention is assumed):
\begin{equation}\label{polarization bivector}
\mathcal{P}^{(\alpha)}(\mathcal{F},\mathcal{F}^\dagger)=-{\epsilon_0\over c}\int_0^\infty\kappa^{(\alpha)}_{mn}(\gamma_m\gamma_n\bol{\cdot}\mathcal{E})_{\chi_0-\xi_0}d\xi_0
\end{equation}
 
Utilizing \eqref{diff2}, \eqref{binabla} and \eqref{conj_diff} one obtains:

\begin{equation}\label{matrix form 2}
\begin{bmatrix}
\mathcal{D}&0\\
0&\mathcal{D}^\dagger
\end{bmatrix}=
\gamma_0\begin{bmatrix}
\partial^0+\overleftrightarrow{\nabla}&0\\
0&\partial^0-\overleftrightarrow{\nabla}
\end{bmatrix}
=
\gamma_0\partial^0
+\begin{bmatrix}
1&0\\
0&-1
\end{bmatrix}\gamma_l\partial^l
\end{equation}
This enables one to re-write \eqref{matrix form 1} as follows:
\begin{equation}\label{matrix form 3}
\partial^0\gamma_0\begin{bmatrix}\mathcal{F}\\ \mathcal{F}^\dagger\end{bmatrix}=
\begin{bmatrix}
-\gamma_l\partial^l&0\\
0&\gamma_l\partial^l
\end{bmatrix}\begin{bmatrix}\mathcal{F}\\ \mathcal{F}^\dagger\end{bmatrix}-
\partial^0\Sigma_\alpha {\gamma_0\over\epsilon_0}\begin{bmatrix} \mathcal{P}^{(\alpha)}\\ \mathcal{P}^{(\alpha)}\end{bmatrix}+\Sigma_\alpha {1\over\epsilon_0}\begin{bmatrix}
\partial^l\gamma_l\bol{\cdot}\mathcal{P}^{(\alpha)}&0\\
0&-\partial^l\gamma_l\bol{\cdot}\mathcal{P}^{(\alpha)}
\end{bmatrix}\
\end{equation}

Projecting the last equation onto the time axis, that is, geometrically multiplying it by $\gamma_0$, yields the following evolution equation (summation convention is assumed for $l=1,2,3$):
\begin{equation}\label{matrix form 3}
\partial^0\begin{bmatrix}\mathcal{F}\\ \mathcal{F}^\dagger\end{bmatrix}=
\begin{bmatrix}
\gamma_{l0}\partial^l&0\\
0&\gamma_{0l}\partial^l
\end{bmatrix}\begin{bmatrix}\mathcal{F}\\ \mathcal{F}^\dagger\end{bmatrix}-
\Sigma_\alpha {1\over\epsilon_0}\begin{bmatrix} \mathcal{J}^{(\alpha)}\\ \mathcal{J}^{(\alpha)}\end{bmatrix}+\Sigma_\alpha {1\over\epsilon_0}
\begin{bmatrix}
\partial^l\gamma_{0l}&0\\
0&\partial^l\gamma_{l0}
\end{bmatrix}
\bol{\cdot}
\begin{bmatrix} \mathcal{P}^{(\alpha)}\\ \mathcal{P}^{(\alpha)}\end{bmatrix}
\end{equation}
with,
\begin{equation}\label{space-time blades}
\gamma_{0l}=\gamma_0\gamma_l,\quad\gamma_{l0}=\gamma_l\gamma_0\end{equation}
and 
\begin{equation}\label{polarization current}
\mathcal{J}^{(\alpha)}\equiv\partial^0\mathcal{P}^{(\alpha)}\end{equation}
is the respective $\emph{polarization current bivector}$ for the species $\alpha$, namely the temporal rate (in spatial units, $m^{-1}$) of the polarization bivector.

This equation points towards the necessity for introducing auxiliary bivectors, the instantaneous state of which must be known, that is, evolving along with the EM bivector and its conjugate. In the following, this will eventually sum up to an augmented state treatment of our problem. 

In the frequency domain, expressions for $\kappa^\alpha_{mn}$ denoted as $\kappa^\alpha_{mn,\omega}(\{\chi_p\},\chi_0)$, one has: 
\begin{equation}\label{frequency domain}
\kappa^{(\alpha)}_{mn}(\{\chi_p\},\chi_0)=\int \frac{d\omega}{2\pi}\kappa^{(\alpha)}_{mn,\omega}(\{\chi_p\})\exp \left(-i\frac{\omega}{c}\chi_0\right)
\end{equation}
For lossless cold magnetized plasma the dimensionless susceptibility matrix $\kappa^{(\alpha)}_{mn,\omega}(\{\chi_p\},\chi_0)$ is \cite{Stix}:
\begin{equation}\label{Stix 1}
\kappa^{(\alpha)}_{mn,\omega}(\{\chi_p\},\chi_0)
=\begin{bmatrix}
\frac{L_\alpha+R_\alpha}{2}&i\frac{L_\alpha-R_\alpha}{2}&0\\
i\frac{R_\alpha-L_\alpha}{2}&\frac{L_\alpha+R_\alpha}{2}&0\\
0&0&P_\alpha
\end{bmatrix}    
\end{equation}
where
\begin{equation}\label{Stix 2}
 L_\alpha=\frac{P_\alpha}{1-\epsilon_\alpha C_\alpha},   
\quad
 R_\alpha=\frac{P_\alpha}{1+\epsilon_\alpha C_\alpha},  \quad
 P_\alpha=-\Big(\frac{\omega_{p\alpha}}{\omega}\Big)^2,  
\quad
 C_\alpha=-\frac{\omega_{c\alpha}}{\omega},
 \quad
 \epsilon_\alpha=\pm{1}
\end{equation}

In the latter, $\omega_{p\alpha}, \omega_{c\alpha}$ and $\epsilon_\alpha$ are respectively the species plasma frequency, gyrofrequency and sign of its charge. In our treatment we allow the plasma frequency (dependent on the species density) to be constant but different in neighboring regions. Note that the susceptibility matrix in the time domain is a skew symmetric matrix reflecting the gyrotropic nature of the magnetized plasma.  
On the other hand, in the frequency domain, the susceptibility matrix $\kappa^{(\alpha)}_{mn,\omega}(\{\chi_p\},\chi_0)$ in \eqref{Stix 1} is  anti-Hermitian. However it must be viewed and accordingly adjusted to fit within the realm of the geometric algebra $\mathcal{DA}$: In \eqref{frequency domain} the exponential must be interpreted as a $\emph{rotor}$. that is, the imaginary unit is actually the pseudoscalar $\mathcal{I}$ (\eqref{pseudoscalar}) introduced in \eqref{Introduction}. On the other hand, the rotation that characterizes the gyrotropic character of our medium is evident via the off-diagonal elements in \eqref{Stix 1}. Thus, we adjust the latter accordingly:
\begin{equation}\label{Stix 2}
\kappa^{(\alpha)}_{mn,\omega}(\{\chi_p\},\chi_0)
=\begin{bmatrix}
\frac{L_\alpha+R_\alpha}{2}&\mathcal{I}\frac{L_\alpha-R_\alpha}{2}&0\\
\mathcal{I}\frac{R_\alpha-L_\alpha}{2}&\frac{L_\alpha+R_\alpha}{2}&0\\
0&0&P_\alpha
\end{bmatrix}    
\end{equation}
Therefore, the geometric representation of the polarization bivector $\mathcal{P}^{(\alpha)}$ in \eqref{polarization bivector} is as follows:

\begin{equation}\label{polarization bivector 2}
\mathcal{P}^{(\alpha)}=
-{\epsilon_0 \over c}\int_0^\infty d\xi_0\int \frac{d\omega}{2\pi}\int_{-\infty}^{\chi_0} d\xi_0\kappa^{(\alpha)}_{mn,\omega}(\{\chi_p\})\exp \left[-\mathcal{I}\frac{\omega}{c}(\chi_0-\xi_0)\right]\left(\gamma_m\gamma_n\bol{\cdot}\mathcal{E}\right)_{\xi_0}
\end{equation}
Note that in this expression, as well as in subsequent similar situations, double summation convention over $m,n=1,2,3$ is assumed. 

The polarization currents can be derived by taking the temporal derivative (in spatial units) of the  RHS of \eqref{polarization bivector 2} as well as \eqref{E and B bivectors}. This yields a ballistic term, which vanishes in the long $\chi_0$ and the non-zero contribution:
\begin{equation}\label{evolution of A bivector}
\mathcal{J}^{(\alpha)}=
\mathcal{I}{\epsilon_0 \over 2c^2}\int_0^\infty d\xi_0\int \frac{d\omega \omega}{2\pi}\int_{-\infty}^{\chi_0} d\xi_0\kappa^{(\alpha)}_{mn,\omega}(\{\chi_p\})\exp \left[-\mathcal{I}\frac{\omega}{c}(\chi_0-\xi_0)\right]\gamma_m\gamma_n\bol{\cdot}\left(\mathcal{F}+\mathcal{F}^\dagger\right)_{\xi_0}
\end{equation}
Upon taking into account that \eqref{Stix 1} is an even function of the frequency $\omega$, the temporal evolution (in spatial units) of the bivectors of polarization currents becomes:
\begin{equation}\label{evolution of J bivector}
\partial^0 \mathcal{J}^{(\alpha)}
={\epsilon_0 \over 2c^3}\int_0^\infty d\xi_0\int \frac{d\omega {\omega}^2}{2\pi}\int_{-\infty}^{\chi_0} d\xi_0\kappa^{(\alpha)}_{mn,\omega}(\{\chi_p\})\exp \left[-\mathcal{I}\frac{\omega}{c}(\chi_0-\xi_0)\right]\gamma_m\gamma_n\bol{\cdot}
\left(\mathcal{F}+\mathcal{F}^\dagger\right)_{\xi_0},\end{equation}
Our goal is to express the RHS of \eqref{evolution of J bivector} in terms of the bivectors $\mathcal{F},\mathcal{F}^\dagger,\mathcal{J},\mathcal{J}^\dagger $. Towards this goal, we split the susceptibility into three distinct polarizations: left-handed circular, right-handed circular and parallel (with respect to the externally imposed homogeneous magnetic field):
\begin{subequations}\label{splitting}
\begin{equation}\kappa^{(\alpha)}_{mn,\omega}=\kappa^{(\alpha,L)}_{mn,\omega}+\kappa^{(\alpha,R)}_{mn,\omega}+\kappa^{(\alpha,\parallel)}_{mn,\omega}\end{equation}
\begin{equation}\kappa^{(\alpha,L)}_{mn,\omega}\equiv-\frac{\omega_{pa}^2}{2\omega(\omega-\epsilon_\alpha\omega_{ca})}\begin{bmatrix}
1&\mathcal{I}&0\\
-\mathcal{I}&1&0\\
0&0&0
\end{bmatrix} 
\end{equation}
\begin{equation}\kappa^{(\alpha,R)}_{mn,\omega}\equiv-\frac{\omega_{pa}^2}{2\omega(\omega+\epsilon_\alpha\omega_{ca})}\begin{bmatrix}
1&-\mathcal{I}&0\\
\mathcal{I}&1&0\\
0&0&0
\end{bmatrix} 
\end{equation}
\begin{equation}\kappa^{(\alpha,\parallel)}_{mn,\omega}\equiv-\frac{\omega_{pa}^2}{2\omega^2}
\begin{bmatrix}
0&0&0\\
0&0&0\\
0&0&2
\end{bmatrix} 
\end{equation}
\end{subequations}
This representation reveals the geometrical meaning of the transverse and longitudinal (to the magnetic field) polarizations that the medium imposes: In the transverse plain the $3\times3$ objects, when they operate, induce geometric multiplication with the pseudoscalar, that is, they produce result in the dual space via the duality transform discussed in the \eqref{Appendix A} (\eqref{dual}). In contrast, in the longitudinal direction, only scalars (grade $0$) elements appear. The evolution equation \eqref{matrix form 3} can be re-written as follows (summation convention is assumed for $l=1,2,3$):
\begin{equation}\label{final evolution equations 1}
\begin{split}
\partial^0\begin{bmatrix}\mathcal{F}\\ \mathcal{F}^\dagger\end{bmatrix}
&=\begin{bmatrix}
\gamma_{l0}&0\\
0&\gamma_{0l}
\end{bmatrix}\partial^l\begin{bmatrix}\mathcal{F}\\ \mathcal{F}^\dagger\end{bmatrix} \\
&+{1\over\epsilon_0}
\begin{bmatrix}
\gamma_{0l}&0\\
0&\gamma_{l0}
\end{bmatrix}
\bol{\cdot}\Sigma_\alpha\partial^l\left(\begin{bmatrix}\mathcal{P}^{(\alpha,L)}\\ \mathcal{P}^{(\alpha,L)}\end{bmatrix}+\begin{bmatrix}\mathcal{P}^{(\alpha,R)}\\ \mathcal{P}^{(\alpha,R)}\end{bmatrix}+\begin{bmatrix}\mathcal{P}^{(\alpha,\parallel)}\\ \mathcal{P}^{(\alpha,\parallel)}\end{bmatrix}\right)\\
&-{1\over\epsilon_0}\Sigma_\alpha \left(\begin{bmatrix}\mathcal{J}^{(\alpha,L)}\\ \mathcal{J}^{(\alpha,L)}\end{bmatrix}+\begin{bmatrix}\mathcal{J}^{(\alpha,R)}\\ \mathcal{J}^{(\alpha,R)}\end{bmatrix}+\begin{bmatrix}\mathcal{J}^{(\alpha,\parallel)}\\ \mathcal{J}^{(\alpha,\parallel)}\end{bmatrix}\right)\end{split}\end{equation}

The evolution equations of the polarizations involved in \eqref{final evolution equations 1} are, by definition, the respective polarization currents:
\begin{equation}\label{final evolution equations 2}
\partial^0\mathcal{P}^{(\alpha;L,R,\parallel)}= \mathcal{J}^{(\alpha;L,R,\parallel)}
\end{equation}

After some lengthy, but straightforward, algebraic manipulation we obtain the evolution equations for the respective polarization currents:
\begin{subequations}\label{final evolution equations 3}
\begin{equation}
\partial^0\mathcal{J}^{(\alpha,L)}={\epsilon_\alpha\omega_{c\alpha}\over c}\mathcal{J}^{(\alpha,L)}-\epsilon_0\frac{\omega_{p\alpha}^2}{4c^2}\begin{bmatrix}
1&\mathcal{I}&0\\
-\mathcal{I}&1&0\\
0&0&0
\end{bmatrix}_{m,n}\gamma_m\gamma_n\bol{\cdot}\left(\mathcal{F}+\mathcal{F}^\dagger\right)
\end{equation}
\begin{equation}
\partial^0\mathcal{J}^{(\alpha,R)}=-{\epsilon_\alpha\omega_{c\alpha}\over c}\mathcal{J}^{(\alpha,R)}-\epsilon_0\frac{\omega_{p\alpha}^2}{4c^2}\begin{bmatrix}
1&-\mathcal{I}&0\\
\mathcal{I}&1&0\\
0&0&0
\end{bmatrix}_{m,n}\gamma_m\gamma_n\bol{\cdot}\left(\mathcal{F}+\mathcal{F}^\dagger\right)
\end{equation}
\begin{equation}
\partial^0\mathcal{J}^{(\alpha,\parallel)}=-\epsilon_0\frac{\omega_{p\alpha}^2}{2c^2}\begin{bmatrix}
0&0&0\\
0&0&0\\
0&0&1
\end{bmatrix}_{m,n}\gamma_m\gamma_n\bol{\cdot}\left(\mathcal{F}+\mathcal{F}^\dagger\right)
\end{equation}
\end{subequations}
 Again, note that double summation convention has been assumed for $m,n=1,2,3$. The grade $2$ spatial blades $\gamma_m\gamma_n, n=1,2,3$ are been added according to the elements of the respective $3\times 3$ matrices on the left that contain scalar and pseudoscalar elements.  The outcome will be a bivector and a scalar. The latter form inner products, from the left, with the EM bivectors leading to bivectors that dictate the evolution of the polarization currents. Thus, utilizing the definition of the pseudo-scalar $\mathcal{I}$ in \eqref{basis}, one readily obtain:
\begin{subequations}\label{final evolution equations 3a}
\begin{equation}
\partial^0\mathcal{J}^{(\alpha,L)}={\epsilon_\alpha\omega_{c\alpha}\over c}\mathcal{J}^{(\alpha,L)}+\mathcal{S}^{(\alpha,L)}\bol{\cdot}\left(\mathcal{F}+\mathcal{F}^\dagger\right)
\end{equation}
\begin{equation}
\partial^0\mathcal{J}^{(\alpha,R)}=-{\epsilon_\alpha\omega_{c\alpha}\over c}\mathcal{J}^{(\alpha,R)}+\mathcal{S}^{(\alpha,R)}\bol{\cdot}\left(\mathcal{F}+\mathcal{F}^\dagger\right)
\end{equation}
\begin{equation}
\partial^0\mathcal{J}^{(\alpha,\parallel)}=\mathcal{S}^{(\alpha,\parallel)}\bol{\cdot}\left(\mathcal{F}+\mathcal{F}^\dagger\right)
\end{equation}
\end{subequations}
where, 
\begin{subequations}\label{S}
\begin{equation}
\mathcal{S}^{(\alpha,L)}=-\epsilon_0\frac{\omega_{p\alpha}^2}{4c^2}\left(\gamma_1\gamma_1 + \gamma_2\gamma_2 -2\gamma_0\gamma_3\right)
\end{equation}
\begin{equation}
\mathcal{S}^{(\alpha,R)}=-\epsilon_0\frac{\omega_{p\alpha}^2}{4c^2}\left(\gamma_1\gamma_1 + \gamma_2\gamma_2 +2\gamma_0\gamma_3\right)\end{equation}
\begin{equation}
\mathcal{S}^{(\alpha,\parallel)}=-\epsilon_0\frac{\omega_{p\alpha}^2}{2c^2}\gamma_3\gamma_3
\end{equation}
\end{subequations}

All the bivectorial entities $\mathcal{F}$, $\mathcal{F}^\dagger$, $\mathcal{P}^{(\alpha;L,R,\parallel)}$ and $\mathcal{J}^{(\alpha,;L,R,\parallel)}$ constitute a columnar state function with $2+6N_\alpha$ elements, where $N_\alpha$ is the number of species involved in the plasma (electrons and various ions). For the sake of transparency and simplicity in the notations, especially for the  Clifford conjugation, we adopt the Dirac's "ket" column-wise notation for this state function:
\begin{equation}\label{state vector}
\ket\Psi\equiv\begin{bmatrix}
\mathcal{F}\\\mathcal{F}^\dagger\\\left(\alpha;{\mathcal{P}^{L}\over\epsilon_0}\right)\\\left(\alpha;{\mathcal{P}^{R}\over\epsilon_0}\right)\\\left(\alpha;{\mathcal{P}^{\parallel}\over\epsilon_0}\right)\\\left(\alpha;{\mathcal{J}^{L}\over\epsilon_0}\right)\\\left(\alpha;{\mathcal{J}^{R}\over\epsilon_0}\right)\\\left(\alpha;{\mathcal{J}^{\parallel}\over\epsilon_0}\right)
\end{bmatrix}
\end{equation}
where $(\alpha;-)$ denotes a sub-column with $N_\alpha$ rows. Note that the respective sub-columns which correspond to the the bivectors $\mathcal{F}$ and $\mathcal{F}^\dagger$ are single-row, while the other sub-columns involve the collective evolution of all plasma species.. This way one is able to obtain the evolution operator in a square matrix form. Now, the evolution equations, \eqref{final evolution equations 1}, \eqref{final evolution equations 2} and \eqref{final evolution equations 3a} can now be written in a compact form:
\begin{equation}\label{compact}
\partial_0\ket\Psi=\mathcal{W}\cdot\ket{\Psi}\end{equation}
The "$\cdot$" on the right of $\mathcal{W}$ signifies operation of the latter on the state from its left.

The $(6N_\alpha+2)\times (6N_\alpha+2)$ operator $\mathcal{W}$ is defined as follows (summation convension is assumed for $r=1,2,3$):
\begin{equation}\label{evolution operator} 
\begin{split}
&\mathcal{W}\equiv \\
&\begin{bmatrix}
\partial^l\gamma_{l0} &0&\left(\partial^l\gamma_{0l}\bol{\cdot}, \alpha\right)&\left(\partial^l\gamma_{0l}\bol{\cdot}, \alpha\right)&\left(\partial^l\gamma_{0l}\bol{\cdot}, \alpha\right)&\left(-1,\alpha\right)&\left(-1,\alpha\right)&\left(-1,\alpha\right)\\
0&\partial^l\gamma_{0l}&\left(\partial^l\gamma_{l0}\bol{\cdot}, \alpha\right)&\left(\partial^l\gamma_{l0}\bol{\cdot}, \alpha\right)&\left(\partial^l\gamma_{l0}\bol{\cdot}, \alpha\right)&\left(-1,\alpha\right)&\left(-1,\alpha\right)&\left(-1,\alpha
\right)\\
\left(0,\alpha\right)&\left(0,\alpha\right)&\left(0,\alpha\right)&\left(0,\alpha\right)&\left(0,\alpha\right)&\left(1,\alpha\right)&\left(0,\alpha\right)&\left(0,\alpha
\right)\\
\left(0,\alpha\right)&\left(0,\alpha\right)&\left(0,\alpha\right)&\left(0,\alpha\right)&\left(0,\alpha\right)&\left(0,\alpha\right)&\left(1,\alpha\right)&\left(0,\alpha
\right)\\
\left(0,\alpha\right)&\left(0,\alpha\right)&\left(0,\alpha\right)&\left(0,\alpha\right)&\left(0,\alpha\right)&\left(0,\alpha\right)&\left(0,\alpha\right)&\left(1,\alpha
\right)\\
\left(\mathcal{S}^{(\alpha,L)}\bol{\cdot},\alpha\right)&\left(\mathcal{S}^{(\alpha,L)}\bol{\cdot},\alpha\right)&\left(0,\alpha\right)&\left(0,\alpha\right)&\left(0,\alpha\right)&\left({\epsilon_\alpha\omega_{c\alpha}\over c},\alpha\right)&\left(0,\alpha\right)&\left(0,\alpha
\right)\\
\left(\mathcal{S}^{(\alpha,R)}\bol{\cdot},\alpha\right)&\left(\mathcal{S}^{(\alpha,R)}\bol{\cdot},\alpha\right)&\left(0,\alpha\right)&\left(0,\alpha\right)&\left(0,\alpha\right)&\left(0,\alpha\right)&\left(-{\epsilon_\alpha\omega_{c\alpha}\over c},\alpha\right)&\left(0,\alpha
\right)\\
\left(\mathcal{S}^{(\alpha,\parallel)}\bol{\cdot},\alpha\right)&\left(\mathcal{S}^{(\alpha,\parallel)}\bol{\cdot},\alpha\right)&\left(0,\alpha\right)&\left(0,\alpha\right)&\left(0,\alpha\right)&\left(0,\alpha\right)&\left(0,\alpha\right)&\left(0,\alpha\right)
\end{bmatrix}
\end{split}
\end{equation}
In \eqref{evolution operator} the notation $(-,\alpha)$ denotes a sub-row with $N_\alpha$ entries. 

Let us now consider the Clifford conjugate of \eqref{compact}. Since Clifford conjugation involves reversion, the matrix-like form of \eqref{compact} renders operation of $\mathcal{W}$ from the right. Therefore, the Clifford conjugate state assumes the form of a "bra", namely:
\begin{equation}\label{cojugate compact}
\partial_0\bra\Psi=(\mathcal{W}\cdot\ket{\Psi})^\dagger=\bra{\Psi}\cdot\mathcal{W}^{T\dagger}\end{equation}
where $\mathcal{W}^{T\dagger}$ denotes the transpose Clifford conjugate of $\mathcal{W}$ and the "$\cdot$" on its left, operation from the right of $\bra\Psi$, the Clifford conjugate of the state vector $\ket\Psi$. The latter consists of the Clifford conjugate sub-columns of $\ket\Psi$ arranged as sub-rows, namely:
\begin{equation}\label{conjugate state vector}
\Psi_{row}^\dagger\equiv
[\mathcal{F}^{\dagger},\mathcal{F},\left(\alpha,{\mathcal{P}^{L}\over\epsilon_0}\right),\left(\alpha,{\mathcal{P}^{R}\over\epsilon_0}\right),\left(\alpha,{\mathcal{P}^{\parallel}\over\epsilon_0}\right),\left(\alpha,{\mathcal{J}^{L}\over\epsilon_0}\right),\left(\alpha,{\mathcal{J}^{R}\over\epsilon_0}\right),\left(\alpha,{\mathcal{J}^{\parallel}\over\epsilon_0}\right)]
\end{equation} 
By simple inspection, one can easily deduce that:
\begin{equation}\label{conjugate compact}
\mathcal{W}^\dagger=\mathcal{W}\end{equation}
Therefore \eqref{conjugate compact} can also be written as follows:
\begin{equation}\label{cojugate compact 2}
\partial_0\bra\Psi=\bra{\Psi}\cdot\mathcal{W}^{T}\end{equation}
It is straightforward, within the geometric product $\bra{\Psi}\ket{\Psi}$, to observe the following scalars directly obtained  from the geometric products among sub-row entities with their respective sub-column ones:
\begin{equation}\label{geometric product for polarizations}
\left(\alpha,{\mathcal{P}^{L,R,\parallel}\over\epsilon_0}\right)\left(\alpha;{\mathcal{P}^{L,R,\parallel}\over\epsilon_0}\right)={1 \over \epsilon^2_0}\Sigma_\alpha P^{(\alpha,L,R,\parallel)}_mP^{(\alpha,L,R,\parallel)}_m={1 \over \epsilon^2_0}\Sigma_\alpha|\bol{P}^{(\alpha,L,R,\parallel)}|^2
\end{equation}
and 
\begin{equation}\label{geometric product for currents}
\left(\alpha,{\mathcal{J}^{L,R,\parallel}\over\epsilon_0}\right)\left(\alpha;{\mathcal{J}^{L,R,\parallel}\over\epsilon_0}\right)={\mu_0 \over \epsilon_0}\Sigma_\alpha J^{(\alpha,L,R,\parallel)}_m J^{(\alpha,L,R,\parallel)}_m={\mu_0 \over \epsilon_0}\Sigma_\alpha|\bol{J}^{(\alpha,L,R,\parallel)}|^2 
\end{equation}
where $\bol{P}^{(\alpha,L,R,\parallel)}$ and $\bol{J}^{(\alpha,L,R,\parallel)}$ are the respective Cartesian vectors for the species $\alpha$. Note that the individual vector components in these triads are mutually orthogonal (the external homogeneous magnetic field is in the z-axial direction).
Combining \eqref{EM energy}, \eqref{geometric product for polarizations} and \eqref{geometric product for currents} one obtains the geometric product of the the state function with its Hermitian conjugate. This boils down to the product of the row-wise Clifford conjugate state function $\bra\Psi$ with its respective column-wise state function $\ket\Psi$:
\begin{equation}\label{state vector norm}
\begin{split}
\bra{\Psi}\ket{\Psi}&={4E_0\over\epsilon_0}
+{1 \over \epsilon^2_0}\Sigma_\alpha\left(|\bol{P}^{(\alpha,L)}|^2+|\bol{P}^{(\alpha,R)}|^2+|\bol{P}^{(\alpha,\parallel)}|^2\right)\\
&+{\mu_0 \over \epsilon_0}\Sigma_\alpha\left(|\bol{J}^{(\alpha,L)}|^2+|\bol{J}^{(\alpha,R)}|^2+|\bol{J}^{(\alpha,\parallel)}|^2\right)\\
&={4E_0\over\epsilon_0}
+{1 \over \epsilon^2_0}\Sigma_\alpha|\bol{P}^{(\alpha)}|^2
+{\mu_0 \over \epsilon_0}\Sigma_\alpha(|\bol{J}^{(\alpha)}|^2
\end{split}
\end{equation}
Note that the polarization currents $L,R,\parallel$ obey their individual conservation laws \eqref{polarization charge conservation} by their own definition \eqref{final evolution equations 2}.
This is a scalar positive definite entity which represents the squared norm of the state function. It is clearly a scalar function of space and time. The most important property of the norm can be derived from \eqref{compact} and \eqref{cojugate compact 2}. It is the following:
\begin{equation}\label{temporal rate of norm}
\partial^0\bra{\Psi}\ket{\Psi} = \bra{\Psi}\cdot\mathcal{W}^T\ket{\Psi}=\bra\Psi\mathcal{W}\cdot\ket{\Psi}
\end{equation}
Here, we have used the notion of the inner product introduced as the symmetrized geometric product of two bivectors (see \eqref{EM energy}). One may identify the squared norm with an \emph{extended EM energy density} that additively contains the $\emph{vacuum energy density}$, the $\emph{polarization energy density}$, as well as the $\emph{polarization momentum flux}$ (flux: per unit time, per unit surface of incidence):
\begin{equation}\label{extended energy}
E_{ext} \equiv  {\epsilon_0\over 4}\bra{\Psi}\ket{\Psi}
=E_0+{1 \over 4\epsilon_0}\Sigma_\alpha|\bol{P}^{(\alpha)}|^2
+{\mu_0 \over 4}\Sigma_\alpha|\bol{J}^{(\alpha)}|^2
\end{equation}
Under suitable boundary conditions the integral of this expression in the domain of interest, $\Omega$, can be initially set to zero. That suffices to render the spatially integrated norm time independent:  
\begin{equation}\label{conservation}
{d\over dt}\int_\Omega E_{ext}d\xi_1d\xi_2d\xi_3 =0  
\end{equation}

It is of tantamount importance to note that the resulting evolution equation and the underlying conservation law distinguish among the three polarizations (two transverse to the externally applied magnetic field and one along the latter) their respective currents in contrast to \cite{Koukoutsis2}. This is expected from the kind of problem the latter is focused on: The evolution of an EM mode that enters into play with the plasma being in the vacuum state. In the present work, the medium in hand is already in a fully developed state, that is, there exist, at least, background EM fluctuations and modes conforming with the magnetized plasma dispersion relation (cold, in the present approach).

One may solve the evolution equation \eqref{compact} in a particular domain in a magnetized plasma. The problem will be treated as an initial value problem ($t=0$) with proper boundary conditions imposed on the domain's boundary. Since our goal is to investigate the propagation of EM waves in the plasma at a state where there exist an already developed mode or a turbulent state (at every internal point in the domain, as well as on the boundary) the plasma will necessarily be initially polarized for a given initial value of the EM bivector $\mathcal{F}$. As far as the boundary is concerned, one can straightforwardly (though, not easily), in a numerical scheme to ensure the magnetic induction as well as the electric displacement to be zero at the boundary. In principle, these requirements can be achieved by introducing proper fictitious sources on the boundary. Thus, the polarization and the polarization energy density in \eqref{extended energy} will not be zero in the interior nor on the boundary. On the other hand, the polarization momentum flux could be set to zero if one consider an adiabatic turning-on of an EM beam that penetrates the boundary. Depending on the particular problem at hand, the initial value of the extended energy through-out the computational mesh, that sets the domain of interest in the three dimensional space (a lattice), will be an independent of time spatial function.  

\section{Conclusion}\label{Conclusion}
We utilized the advantages of Clifford Algebras to reformulate the propagation of electromagnetic waves in a cold multi-species lossless magnetized plasma in a state where a global mode has been established or there exist electromagnetic fluctuations. The goal of this work is to provide a numerically implementable theoretical tool in the realm of complex initial and boundary conditions that characterize the entirety of applications in the thermonuclear fusion research. Clifford algebra provides a coordinate free point of view and it is especially capable to handle symmetries and to "uncover" conservation laws (like energy) in a complex medium such as a multi-species magnetized plasma.

The formulation led to a real Dirac type evolution equation for an augmented state that consists of the electric and magnetic field bivector and its conjugate, as well as the polarizations and their associated currents for each species and the three distinct polarizations that characterize the propagation in a magnetized plasma. This evolution equation along with the evolution equation of the Clifford conjugate augmented state can be dealt with by a general spatial lattice disretization scheme. The evolution operator involved coincides with its Clifford conjugate. That is, the temporal advancement of the state is Hermitian (in the realm of Clifford algebra). An important quantity is the extended energy density, the sum of the vacuum field energy density, the polarization density and the polarization momentum flux. Depending on the particular problem at hand, the initial value of the extended energy (integrated extended energy density) through-out the computational mesh, that sets the domain of interest in the three dimensional space (a lattice), will be an independent of time spatial function. The formulation becomes computationally simpler whatever the application could be and suitable for quantum computation. The latter could be ideal if we are seeking small wavelength applications (on the Debye length scale) and spatially large systems (magnetic confinement devices) where a numerical solution for conventional classical super-computer becomes questionable.  In the Appendix B, it is shown how one may apply the present work in a rudimentary quantum computer.

\appendix
\section{Basics of $\mathcal{C}l({\mathbb{R}}^{1,3})$}\label{Appendix A}
The Dirac Algebra, $\mathcal{DA}$ is a remarkably powerful tool that describes scalars along with 4-dimensional objects such as vectors, planes, volumes and pseudo-scalars. It contains all of the familiar vector operations, but most importantly a new type of algebraic product, the so-called Geometric or Clifford product. For vectors, denoted by bold face letters, $\boldsymbol{a},\boldsymbol{b}\in{\mathbb{R}}^3$
\begin{equation}\label{geometric product 1}
\boldsymbol{a}\boldsymbol{b}=\boldsymbol{a}\bol{\cdot}\boldsymbol{b}+\boldsymbol{a}\wedge\boldsymbol{b}
\end{equation}
The result is the sum of a scalar (the so-called inner product) and the so-called wedge ($\wedge$) or exterior product which is called \emph{bivector}. Thus, it produces the sum of two distinct objects that forms a \emph{multivector} just like the sum in the complex numbers. Orthonormal vectors render, by definition, zero inner product. In the Dirac Algebra we choose an orthonormal basis of vectors denoted with $\{\gamma_\nu, \nu=0,1,2,3\}$, or, equivalently, with $\{\gamma_0,\gamma_m, m=1,2,3\}$ that satisfy the following generalized orthonormality conditions:
\begin{subequations}\label{orthonormality}
\begin{equation}
\gamma_0\gamma_0=1,\quad \gamma_m\bol{\cdot}\gamma_n=-\delta_{mn},\quad \gamma_0\bol{\cdot}\gamma_m=0, \quad m,n=1,2,3,
\end{equation} 
as well as:
\begin{equation}
 \gamma_\mu\wedge\gamma_\mu=0,\quad \mu=0,1,2,3,
\end{equation} 
\end{subequations}
 
The ${\mathbb{R}}^{1,3}$ space now has been equipped with the product of Equation \eqref{2.1}, hence the basis set of the  $\mathcal{DA}$ is generated from the orthonormal basis $\{ \gamma_\mu\}$ vectors, rendering five different bases for the respective five geometrical grades, namely the scalars and 4-dimensional (spacial and temporal)  vectors, the 4-dimensional bivectors (space-space and space-time planes or blades), the trivectors (spatial oriented volumes and spacetime oriented volumes) and the pseudo-scalars (the highest grade geometrical element). Furthermore, because of \eqref{orthonormality} one can actually suppress the exterior product in the process of defining the bases for the various grades:
\begin{subequations}\label{basis}
\begin{equation}
\{1\}
\end{equation}
\begin{equation}
\{\gamma_0,\gamma_1,\gamma_2,\gamma_3\}
\end{equation}
\begin{equation}
\{\gamma_1\gamma_0,\gamma_2\gamma_0, \gamma_3\gamma_0,\gamma_1\gamma_2 ,\gamma_2\gamma_3,\gamma_3\gamma_1\}
\end{equation}
\begin{equation}
\{\gamma_1\gamma_2\gamma_3,\gamma_1\gamma_2\gamma_0, \gamma_2\gamma_3\gamma_0, \gamma_3\gamma_1\gamma_0\}
\end{equation}
\begin{equation}
\{\gamma_0\gamma_1\gamma_2\gamma_3\}
\end{equation}
\end{subequations} 
The latter basis element is usually denoted as $\mathcal{I}=\gamma_0\gamma_1\gamma_2\gamma_3$. All the elements of $\mathcal{DA}$ can be decomposed in elements (multivectors) that belong to different grades. On the other hand, the pseudoscalar has a paramount intrinsic geometrical significance: It commutes with elements of even grade and anti-commutes with elements of odd grade. If $\mathcal{A}_r$ is a multivector of grade $r$ (a pure multivector), then:
\begin{equation}\label{pseudo_blade_mul}
\mathcal{A}_r\mathcal{I}=(-1)^{3r}\mathcal{I}\mathcal{A}_r
\end{equation}
As a consequence of the choice of the Minkowskian metric $(1,-1,-1,-1)$, the geometric product of the pseudo-scalar with itself (its square) is scalar and:
\begin{equation}\label{pseudo_square}
\mathcal{I}\mathcal{I}={\mathcal{I}}^2=-1
\end{equation}
The LHS of \eqref{pseudo_blade_mul} is the (minus) so called dual (prefix $\star$) of the pure blade $\mathcal{A}_r$, or, equivalently, its orthogonal complement (superscript $\perp$). Generally speaking, going from a multivector to a multivector via multiplication by the pseudoscalar is the so-called $\emph{duality transformation}$:
\begin{equation}\label{dual}
\star\mathcal{A}_r\equiv{\mathcal{A}_r}^\perp\equiv-\mathcal{A}_r\mathcal{I}
\end{equation}
The orthogonal complement of a pure blade of grade $r$ does not contain vectors that “lie on” the blade, since the grade of the orthogonal complement is $4-r$. For bivectors, this amounts to going (for example) from space-time bivectors to purely spatial ones. 

The duality transform is of tantamount importance in reformulating Maxwell equation to a Schr\"odinger-Dirac form. In the framework of $\mathcal{DA}$, the proper “tools” are (1) the so-called $\emph{grade involution}$, specifically called $\emph{space conjugation}$: It is the geometric multiplication of an object $\mathcal{A}$ from both sides by the time-like direction $\gamma_0$ (superscript $\star$):
\begin{equation}\label{grade involution}
{\mathcal{A}}^\star\equiv\gamma_0\mathcal{A}\gamma_0,
\end{equation}
and (2) $\emph{reversion}$ (over $\sim$). In terms of a geometric product of grade one objects (vectors) this can readily defined as follows:
\begin{equation}\label{reversion1}
\widetilde{(abc...z)}\equiv(z...cba)
\end{equation}
This amounts to the following change of signs (from plus) in the separation of grades (signified by ${\langle\rangle}_r, r=0...4$) in an object $\mathcal{A}$ of $\mathcal{DA}$:
\begin{equation}\label{reversion2}
    \mathcal{A} = \langle \mathcal{A}\rangle_0+\langle \mathcal{A}\rangle_1+\langle \mathcal{A}\rangle_2+\langle \mathcal{A}\rangle_3+\langle \mathcal{A}\rangle_4,\quad\widetilde{\mathcal{A}}=\langle \mathcal{A}\rangle_0+\langle \mathcal{A}\rangle_1-\langle \mathcal{A}\rangle_2-\langle \mathcal{A}\rangle_3+\langle \mathcal{A}\rangle_4
\end{equation}
Both involutions (grade and reversion) commute by definition. The application of both is called Clifford conjugation ($\dagger$) in $\mathcal{DA}$. It is an extension of Hermitian conjugation from the Pauli Algebra $\mathcal{PA}$ to the Dirac Algebra $\mathcal{DA}$ (that is why the same symbol is used, although there are different symbolisms in the literature). It is also called relative (relative to the time-like direction $\gamma_0$) reversion-involution:
\begin{equation}\label{conjugation}
\mathcal{A}^\dagger\equiv\gamma_0\widetilde{\mathcal{A}}\gamma_0=\widetilde{\mathcal{A}}^\star
\end{equation}
As far as the EM bivector is concerned, it is straightforward to obtain its reversion as well as its Clifford conjugate :
\begin{equation}\label{EM bivector properties}
\tilde{\mathcal{F}}=-\mathcal{F}, \quad \mathcal{F}^\dagger=\mathcal{E}-\mathcal{I}\mathcal{B}
\end{equation}
It is also important in our application to generalize the geometric product \eqref{geometric product 1} to the one between a vector or vector operator $\boldsymbol{a}$ (grade $1$ objects) and a bivector $\mathcal{B}$:
\begin{equation}\label{geometric product 2}
\boldsymbol{a}\mathcal{B}=\boldsymbol{a}\bol{\cdot}\mathcal{B}+\boldsymbol{a}\wedge\mathcal{B}
\end{equation}
Above, both the vector and the bivector are expressed in the basis \eqref{basis} and the inner and exterior products are executed in accordance with \eqref{orthonormality}.

We now introduce the Dirac differential operator. It is a grade $1$ (vectorial) differential operator and it involves the 4-tangent vectors in the Minkowskian space and the partial differentiation of the differentiated object of $\mathcal{DA}$. Its form is (summation convention is adapted):
\begin{equation}\label{diff1}
    \mathcal{D}\equiv\gamma_\mu\partial^\mu,\quad\mu=0,1,2,3
\end{equation}
Seeing this gradient operator as a four-vector operator, operating in $\mathcal{DA}$, one can "split" it by its right geometric product with the time-like direction $\gamma_0$ as follows (summation convention is adapted):
\begin{equation}\label{split}
\mathcal{D}\gamma_0=\mathcal{D}\bol{\cdot}\gamma_0+\mathcal{D}\wedge\gamma_0=\partial^0+\gamma_m\wedge\gamma_0\partial^m=\partial^0+\gamma_m\gamma_0\partial^m,\quad m=1,2,3
\end{equation}
The second part is clearly a differential bivector acting on each space-time blade $\gamma_m\gamma_0$. We introduce the double arrow to signify this space-time differential operator: 
\begin{equation}\label{binabla}
    \overleftrightarrow{\nabla}\equiv\gamma_0\wedge\gamma_m\partial^m=\gamma_0\gamma_m\partial^m
\end{equation}
Thus, similarly:
\begin{equation}\label{diff2}
    \mathcal{D}=(\partial^0-\overleftrightarrow{\nabla})\gamma_0=\gamma_0(\partial^0+\overleftrightarrow{\nabla})=\tilde{\mathcal{D}},\end{equation}
That is, the time-like direction splits the 4-dimensional gradient into two mutually orthogonal parts: the time-like scalar differential operator and the space-like bivectorial one. The Clifford conjugate of the operator $\mathcal{D}$ can be easily deduced in analogy to the Clifford conjugation of the 4-vectors. Because of \eqref{diff2}:
\begin{equation}\label{conj_diff}
    \mathcal{D}^\dagger=(\partial^0+\overleftrightarrow{\nabla})\gamma_0=\gamma_0(\partial^0-\overleftrightarrow{\nabla})=\tilde{\mathcal{D}^\dagger}={\mathcal{D}}^\star\end{equation}
%%\begin{equation}\label{conj_diff}
%%\widetilde{\mathcal{D}}=\mathcal{D},\quad {\mathcal{D}}^\dagger={\mathcal{D}}^\star
%%\end{equation}
That is, the Clifford (Hermitian) conjugate (or relative reversed) of the 4-dimensional differential operator $\mathcal{D}$ with its space-conjugate. This is something to be expected from the very definition of the relative inversion acting on grade $1$ objects and operators. Notice that both the differential bivector and its Clifford conjugate coincide with their reverse ones.

In Sec.\eqref{Dirac form of Maxwell equations in cold magnetized plasmas} one needs to evaluate the Clifford conjugate of the operation of $\mathcal{D}$ on the EM bivector $\mathcal{F}$:
\begin{equation}\label{conjugate of DF}
(\mathcal{D}\mathcal{F})^\dagger=\gamma_0\tilde{\mathcal{F}}\tilde{\mathcal{D}}\gamma_0=\mathcal{F}^\dagger\mathcal{D}^\dagger=\widetilde{\tilde{\mathcal{D}^\dagger}\tilde{\mathcal{F}^\dagger}}=-\widetilde{\mathcal{D}^\dagger\mathcal{F}^\dagger}\end{equation}
or, equivalently:
\begin{equation}\label{conjugate of DF 2}
\mathcal{D}^\dagger\mathcal{F}^\dagger=-\widetilde{(\mathcal{D}\mathcal{F})^\dagger}
\end{equation}
From \eqref{conjugate Maxwell again} one has to reverse the RHS of this equation and then take its Clifford conjugate. Before we proceed, one must easily notice that the inner product part of the differential operator (a one-vector operator) as well as its Clifford conjugate with a bivector coincides with its reverse:
\begin{equation}\label{reverse of a dot product}
\widetilde{\mathcal{D}^\dagger\bol{\cdot}\mathcal{P}}=\mathcal{D}^\dagger\bol{\cdot}\mathcal{P}
\end{equation}
and,via the relations \eqref{diff2} and \eqref{conj_diff} one obtains:
\begin{equation}\label{conjugate of the reverse of a dot product}
\left(\widetilde{\mathcal{D}^\dagger\bol{\cdot}\mathcal{P}}\right)^\dagger=(\mathcal{D}^\dagger\bol{\cdot}\mathcal{P})^\dagger=-\mathcal{D}\bol{\cdot}\mathcal{P}\end{equation}

\section{An example of computational resources requirement}\label{Appendix B}
Solving Schrödinger/Dirac Equation in the 4-dimentional Minkowskian space, \eqref{compact}, requires spatial discretization. Therefore, there is a need for registering the state at every point of our discretized lattice or network. Thus, one must temporally evolve a rather huge number of EM field data.  Since only bivectors are the elements of interest in the problem in hand, Dirac equation will consist of the 6 independent bivector components in $\{\mathcal{F},\mathcal{F}^\dagger\}$, that is, $\{\gamma_1\wedge\gamma_0,\gamma_2\wedge\gamma_0, \gamma_3\wedge\gamma_0,\gamma_1\wedge\gamma_2 ,\gamma_2\wedge\gamma_3,\gamma_3\wedge\gamma_1\}$ along with the $2\times 3\times 6\times S=36S$ independent bivector components for the S species involved ($3$ counts for $L, R$ and $\parallel$ polarizations and $2$ for the two augmented fields). Therefore, one needs to register $6(6S+1)$ bivector real amplitudes at each register. 

Let us consider $M$ Lattice sites. Let us also consider that we will need $C$ control registers at each lattice side. Therefore, the number of real data needed is $M\times [6(6S+1) +C]$. For example: Consider a Deuterium plasma of electrons and deuterons ($S=2$), confined in a volume $V= 840 m^3$ with electron temperature at the edge (where the plasma can be treated as “cold”) $T = 10^6 K$ (ITER-like parameters). Then, the electron and ion densities $n_{e,i}$ is approximately $10^{17} m^{-3}$ and the Debye length $\lambda_D = 2\times 10^{-4}m$. For an electromagnetic wave to propagate collectively in a plasma the respective wave-length $\lambda$ must be greater than Debye length ($\lambda>\lambda_D$). In that sense the, best resolution that we can ever achieve in a plasma wave propagation simulation is in the Debye scale. To achieve such a resolution, we discretize each spatial dimension. This will lead to $M=V/{\lambda^3_D} \approx 10^{14}$ segments. With C=2, one obtains: 
\begin{equation}\label{volume}
\mathcal{N}=8\times 10^{15}
\end{equation}
This is an extremely demanding number of resources, almost beyond the capabilities of state-of-the art supercomputers. However, one may confine the requirements for computing resources, either by decreasing the volume, or by considering much longer wavelengths for the electromagnetic problem at hand. 

However, in a future quantum qubit implementation \cite{Nielsen} of the Dirac equation one will need:
\begin{equation}\label{qubit number}
\mathcal{N}_q=log_2 \mathcal{N}\approx 53
\end{equation}
Therefore, we need to assign only 53 qubits for our simulation. Most importantly, this requirement is already within the capabilities of the rudimentary quantum computers in our days. 
The aforementioned collection of $\mathcal{N}_q$ qubits can be viewed as multiqubit STGA (MSTGA), that is, the geometric Clifford algebra in Minkowskian space description of N qubit quantum states. The MSTA approach leads to a useful conceptual unification where the states in the multiqubit space and the space of unitary operators in acting on the former, become united.

\bibliography{ref}

\end{document}